\documentclass[conference]{IEEEtran}
\usepackage{cite}
\usepackage{amsmath,amssymb,amsfonts}
\usepackage{algorithmic}
\usepackage{graphicx}
\usepackage{textcomp}
\usepackage{xcolor}
\usepackage{array}
\usepackage{caption}
\usepackage{tabularx,ragged2e,booktabs}
\usepackage{multirow}
\usepackage{multicol}
\usepackage{enumitem}
\usepackage{amsmath}
\usepackage{amssymb}
\usepackage{subfig}
\usepackage{makecell}
\usepackage{float}
\newcolumntype{Z}{ >{\centering\arraybackslash}X }
\usepackage{adjustbox}
\usepackage{listings}
\usepackage{hyperref}
\usepackage{flushend}
\pdfoutput=1

\usepackage[autolanguage, np]{numprint}

\makeatletter
\g@addto@macro{\UrlBreaks}{\UrlOrds}
\makeatother

\begin{document}

\title{Replaying Archived Twitter: When your bird is broken, will it bring you down?
}

\makeatletter
\newcommand{\linebreakand}{%
  \end{@IEEEauthorhalign}
  \hfill\mbox{}\par
  \mbox{}\hfill\begin{@IEEEauthorhalign}
}
\makeatother

\author{
\IEEEauthorblockN{Kritika Garg}
\IEEEauthorblockA{\textit{Department of Computer Science} \\
\textit{Old Dominion University}\\
Norfolk, Virginia, USA \\
\href{mailto:kgarg001@odu.edu}{kgarg001@odu.edu} \\
\href{https://orcid.org/0000-0001-6498-7391}{0000-0001-6498-7391}}
\and
\IEEEauthorblockN{Himarsha R. Jayanetti}
\IEEEauthorblockA{\textit{Department of Computer Science} \\
\textit{Old Dominion University}\\
Norfolk, Virginia, USA \\
\href{mailto:hjaya002@odu.edu}{hjaya002@odu.edu} \\
\href{https://orcid.org/0000-0003-4748-9176}{0000-0003-4748-9176}}
\and
\IEEEauthorblockN{Sawood Alam}
\IEEEauthorblockA{\textit{Wayback Machine} \\
\textit{Internet Archive} \\
San Francisco, California, USA \\
\href{mailto:sawood@archive.org}{sawood@archive.org} \\
\href{https://orcid.org/0000-0002-8267-3326}{0000-0002-8267-3326}}
\linebreakand 
\IEEEauthorblockN{Michele C. Weigle}
\IEEEauthorblockA{\textit{Department of Computer Science} \\
\textit{Old Dominion University}\\
Norfolk, Virginia, USA \\
\href{mailto:mweigle@cs.odu.edu}{mweigle@cs.odu.edu} \\
\href{https://orcid.org/0000-0002-2787-7166}{0000-0002-2787-7166}}
\and
\IEEEauthorblockN{Michael L. Nelson}
\IEEEauthorblockA{\textit{Department of Computer Science} \\
\textit{Old Dominion University}\\
Norfolk, Virginia, USA \\
\href{mailto:mln@cs.odu.edu}{mln@cs.odu.edu} \\
\href{https://orcid.org/0000-0003-3749-8116}{0000-0003-3749-8116}}
}
\maketitle
\begin{abstract}
Historians and researchers trust web archives to preserve social media content that no longer exists on the live web. However, what we see on the live web and how it is replayed in the archive are not always the same. In this paper, we document and analyze the problems in archiving Twitter ever since Twitter forced the use of its new UI in June 2020. Most  web archives were unable to archive the new UI, resulting in archived Twitter pages displaying Twitter's ``Something went wrong” error. The challenges in archiving the new UI forced  web archives to continue using the old UI. To analyze the potential loss of information in web archival data due to this change,  we used the personal Twitter account of the 45\textsuperscript{th} President of the United States, @realDonaldTrump, which was suspended by Twitter on January 8, 2021. Trump’s account was heavily labeled by Twitter for spreading misinformation, however we discovered that there is no evidence in web archives to prove that some of his tweets ever had a label assigned to them. We also studied the possibility of temporal violations in archived versions of the new UI, which may result in the replay of pages that never existed on the live web. Our goal is to educate researchers who may use web archives and caution them when drawing conclusions based on archived Twitter pages. 
\end{abstract}

\section{Introduction}
 Web archives are important digital preservation tools that allow researchers to study the web by replaying its content as it looked in the past on the live web\footnote{The title of this paper is a reference to both Twitter's mascot ``Larry the Bird'' and ``And your bird can sing'' by The Beatles.}. Historians and researchers rely on web archives to preserve the social media records of historical significance and to study these records by replaying them as desired. However, archiving is not a consideration for social media platforms, and existing archives like the Internet Archive's Wayback Machine, archive.today, and perma.cc have difficulties in replaying archived social media content.

One example of this occurred when Twitter shut down its legacy user interface (UI) on June 1, 2020 and forced all users to their new mobile-inspired layout \cite{IntroducingNewTwitter:Twitter}. Unfortunately, Twitter changing its user interface had a detrimental effect on web archives such that they were no longer able to easily archive Twitter at scale. Web archives not being able to capture and replay Twitter pages properly meant that many of the historically significant events that occurred via Twitter in 2020 were not accurately preserved. For example, a historian trying to use archived information in web archives for a study of significant tweets made in late 2020 might witness: a) archived pages, or \emph{mementos} \cite{memento:rfc}, with different user interfaces for the same content, b) mementos displaying the ``Something went wrong” Twitter error message, c) mementos displaying partial content, d) tweets missing from mementos of Twitter account pages, e) mementos not displaying labels and warning messages assigned by Twitter on tweets with disputed information. 

Our goal is to educate and inform researchers looking to utilize this archival content. For this study, we chose former US President Donald Trump’s personal Twitter handle, @realDonaldTrump. We focused on @realDonaldTrump because 
his Twitter activity holds cultural and historical significance and, as one of the most popular Twitter accounts, it is well-archived by multiple sites. Because his account was permanently suspended by Twitter on January 8, 2021 \cite{PermanentSuspensionofTrump:Twitter}, researchers and historians now have to rely on these archived materials for any future studies. This emphasizes how web archives are our primary source of information for social media accounts that have been removed from the live web. 

In this paper, we discuss the challenges faced by web archives in preserving Twitter after the change in Twitter’s UI. Initially, most web archives trying to archive Twitter’s new UI failed. However, we realized that Twitter still uses its old UI for certain user agents, such as ``Googlebot'' \cite{TwitterWasAlreadyDifficult:KritikaHimarsha}. Web archives such as the Internet Archive (IA) and archive.today are preserving Twitter through its old UI by pretending to be another bot. 
However, there is a structural difference between Twitter’s old UI and new UI that causes a gap between the content displayed in archives and what is shown on the live web. Figure~\ref{fig:missing_labels} and Figure~\ref{fig:temporal_range} provide two examples of this gap. 

\begin{figure*}[htbp]\centerline{\includegraphics[width=1\textwidth]{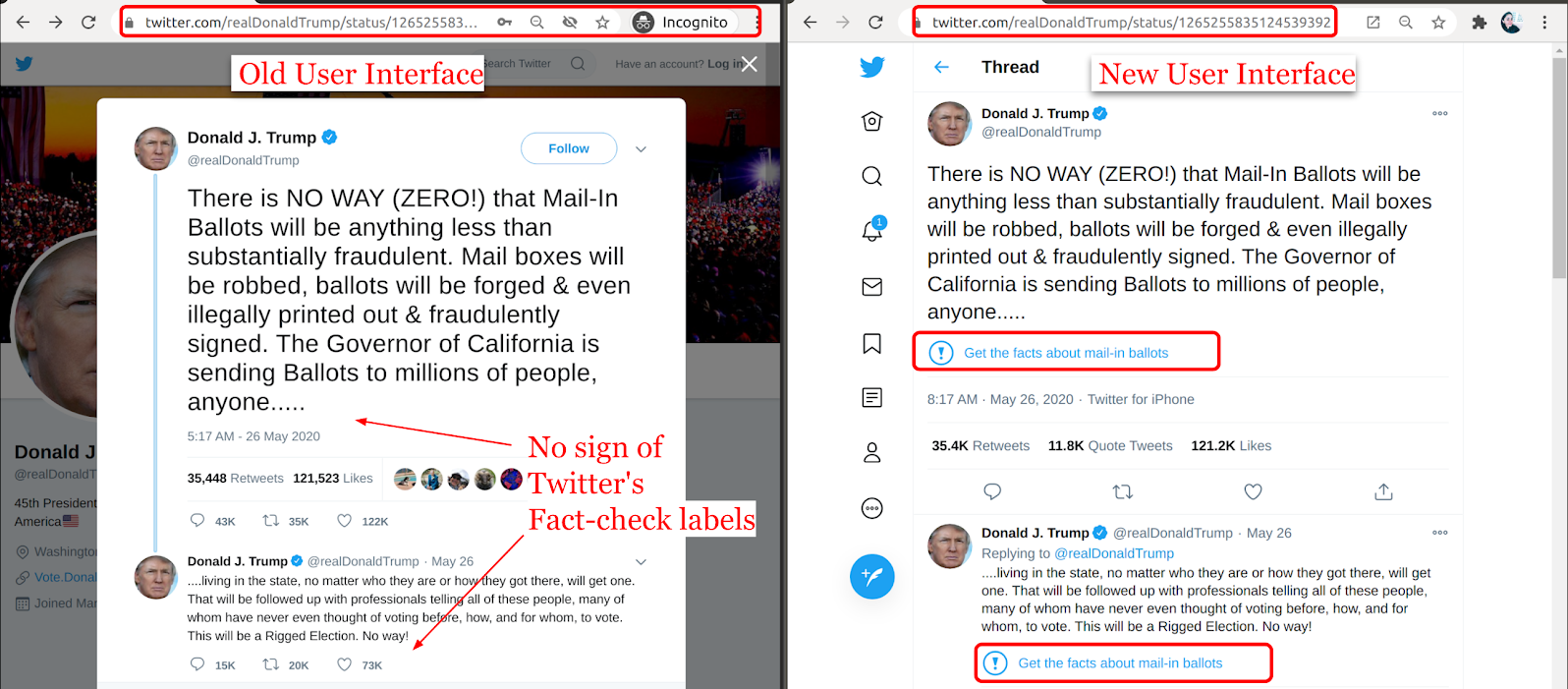}}
\caption{Old UI (left) missing Twitter labels,  and new UI (right) displaying the labels in the live web.}
\label{fig:missing_labels}
\end{figure*}

In May 2020, Twitter introduced new labels to combat misinformation related to COVID-19 and the US presidential election \cite{UpdatingOurApproach:YoelNick}. On May 26, 2020, Twitter added fact-check labels to two of the tweets by @realDonaldTrump related to mail-in voting. On replaying these tweets through web archives, most of the archived pages failed to include the label. Figure~\ref{fig:missing_labels} shows how the labels are missing in Twitter's old UI. This is because the labels were a feature of the new UI, which means that web archives archiving Twitter’s old UI would not have these labels. 
Towards the end of August 2020, we observed that Twitter added the ``Violated Twitter Rules'' (VTR) label to its old UI. In this paper we provide a timeline where different types of labels appeared on the live web but not in Twitter's old UI. In this paper, we assess the potential loss of information in Twitter mementos with respect to labels based on how much of Twitter archived pages are of the old UI. Our second example (Figure~\ref{fig:temporal_range})  focuses on a new UI memento with 71 tweets missing in its tweet feed. On August 18, 2020, @realDonaldTrump's Twitter account page was archived, but the tweets displayed in the memento are from a different day. The most recent tweet in the archived page was from August 16, 2020, almost two days earlier than the archival time, and 71 tweets were missing from the memento. This phenomenon is referred to as a \emph{temporal violation} \cite{ainsworth2014framework}. We discovered that new UI mementos are highly susceptible to temporal violations \cite{NewTwitterUI:HimarshaKritika}. The temporal violation can occur with components from either the past or future and can result in replaying pages that never existed on the live web.

In this paper, we discuss the issues in archived Twitter caused by changes in Twitter’s UI and analyze their occurrence across different web archives. The issues are applicable to other Twitter accounts as well, but we were able to quantify the loss of information with @realDonaldTrump since there are external ground truth datasets for Trump's Twitter account. For other accounts that were suspended after January 6, 2021 \cite{TweetJan6:Twitter}, such as former National Security Advisor Michael Flynn (@GenFlynn) and former federal prosecutor Sidney Powell (@SidneyPowell1), researchers would have to rely on web archives. The web archives would not tell the whole story due to issues such as temporal violation and missing Twitter labels. For example, if researchers want to study Twitter labels, like when was the label added to the tweet, what type of label, and which tweets were labeled using web archives, they should heed the caveats in our paper. After a change in Twitter's UI, studying Twitter by  using only web archives would not give the full picture.

\section{Background/Related Work}
\label{sec:background}
Web archives crawl the web and/or accept individual URL submissions, resulting in  archived pages, or \emph{mementos}. The Memento protocol introduces machine-readable archival metadata, and content negotiation in the dimension of datetime \cite{memento:rfc}. The Memento protocol also provides the following terms: \begin{itemize}
\item URI-R - an original resource from the live Web that exists or used to exist
\item URI-M - an archived version (memento) of a prior state of a URI-R
\item URI-T - a TimeMap that provides a list of mementos (URI-Ms) for a particular URI-R
\item Memento-Datetime - the datetime of when the prior state of an original resource was captured
\end{itemize}
The Memento protocol is supported by most  public web archives, including the Internet Archive (IA). If resources are missing from the live web, web archives allow users to go back in time and access the resource. This is especially useful for social media; for example, in the case of Twitter, we cannot count on the live web to persist due to changes in UI, account suspensions, account closure, 
addition of new labels, etc. 
Since @realDonaldTrump was one of the most popular Twitter accounts, it was regularly archived in various platforms, including third-party databases like thetrumparchive.com\footnote{https://www.thetrumparchive.com/} and factba.se\footnote{https://factba.se/topic/flagged-tweets}. Most accounts, even popular accounts, do not have dedicated third-party datasets. Due to the former president's extensive use of Twitter \cite{wells2020trump}, @realDonaldTrump was a highly active account and has been used in various research studies focusing on his specific tweeting practices \cite{pain2019president} \cite{ott2020twitter} \cite{ExaminingTwitter}. These studies used either thetrumparchive.com or the live web Twitter to collect their data. For suspended Twitter accounts with no third party archive, replicating these studies would require web archives. However, using web archives for this purpose may be confusing due to the issues pointed in our paper.
 In our study, we focused on two types of Twitter pages: a) account/profile pages (Figure~\ref{fig:temporal_range}), and b) tweet/status pages (Figure~\ref{fig:missing_labels}). The account page is displayed when visiting a public Twitter account URI-R, for example, https://twitter.com/realDonaldTrump. Initially, the account page contains 20 tweets in the tweet section, but more tweets can be loaded later by the user. 
A tweet page displays a single tweet (and its replies) and has a URI-R structure of https://twitter.com/realDonaldTrump/status/\{tweet-id\}. We did not include other Twitter pages such as personalized timeline pages and search pages in our study. These pages are likely susceptible to the same problems defined in our study, but they are less likely to be archived/replayed compared to the account and tweet pages.

There is a vast difference in the architecture of Twitter’s old UI and  new UI (Figure~\ref{fig:missing_labels}). In the old  UI, the majority of the content is embedded in the root HTML itself. When a Twitter account page is loaded anonymously (without logging in), the returned root HTML contains 20 embedded tweets, the bio providing a brief description of the account’s owner, and a section for users to sign up or log into Twitter. The HTML also contains empty sections in the sidebar, which are populated with follow-up XHR requests. Successful responses to these requests contain data formatted in JSON that populates the sections Media timeline, ``You might like”, and ``Worldwide trends''/``What's happening''. In the new Twitter UI, all page sections are served dynamically. The root HTML of the new UI contains only a skeleton, and the content is populated later through API JSON responses (tweet JSON) in each of the five sections: Tweet feed, Bio, Media timeline, ``You might like”, and ``What's happening” (Figure \ref{fig:temporal_range}).

To replay the new UI page successfully, web archives must capture the root HTML page along with the asynchronous JSON API responses. This means that to archive the new UI, multiple calls have to be issued to Twitter’s API. However, Twitter  will return an HTTP 429 ``Too Many Requests” response upon a client exceeding its rate limit \cite{Response:Twitter}. To overcome this issue, web archives such as IA and archive.today continued preserving the old UI instead of the new UI. However, the new Twitter UI contains crucial information, such as labels, that the old Twitter UI does not. 

On May 26, 2020, a tweet by @realDonaldTrump was labeled for violating Twitter rules. After then, several tweets by @realDonaldTrump were labeled by Twitter \cite{AdditionalSteps:VijayaKayvon}. We broadly categorize the different types of labels applied by Twitter for tweets with misleading content into ``Fact-check” and VTR labels. Interactions with ``Fact-check” labeled tweets are not disabled even after labeling, whereas VTR labeled tweets were limited to only quote tweeting after the label was applied \cite{TwitterAddedLabels:HimarshaKritika}. The  VTR label was initially a feature of only the new UI and did not exist in the old UI. This label was later added to the old UI, but at the time of this study, the ``Fact-check” labels were not a part of Twitter’s old UI. 

Figure~\ref{fig:labeltimeline} shows the events associated with a labeled tweet on the live web and in web archives:
\begin{itemize}
\item $T_1$: The creation time of the tweet. 
\item $T_2$: The time when Twitter labels the tweet. 
\item $T_3$: The time when a tweet might have been archived by an individual or crawler on IA. 
\item $T_{3-1}$: The tweet is archived before the label is applied to it on the live web. 
\item $T_{3-2}$: The tweet is archived after the label is applied to it on the live web. 
\end{itemize}

\begin{figure*}[htbp]
\centerline{\includegraphics[width=0.85\textwidth]{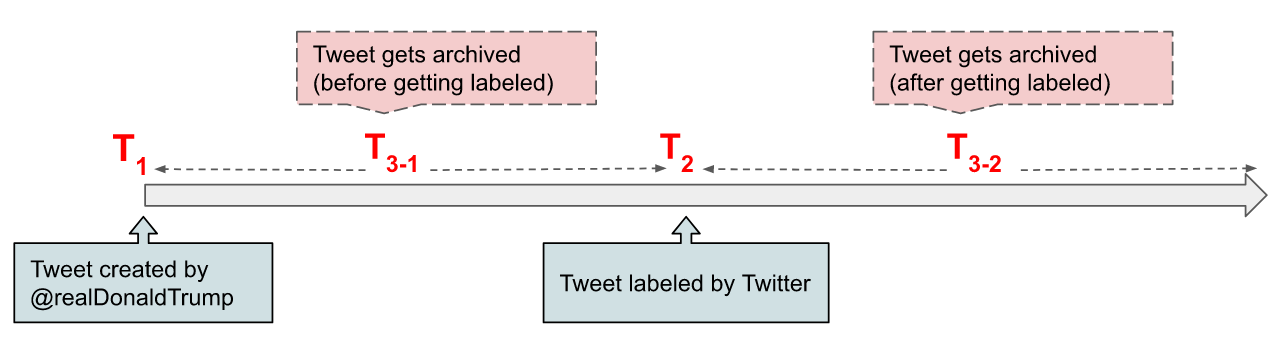}}
\caption{A timeline displaying the two possible windows when the labeled tweets might have been archived.}
\label{fig:labeltimeline}
\end{figure*}

Although there was an average gap of 4--5 hours between $T_1$ and $T_2$ when Twitter first started adding labels, this gap has rapidly closed. By January 2021, before @realDonaldTrump was suspended, Twitter was adding labels almost instantly if misleading content was posted. However, these labels cannot be replayed in web archives that  preserve only the old Twitter UI. In the case of a ``Fact-check” labeled tweet (after $T_2$), anyone can use web archives to prove a tweet was labeled (by using a new UI memento) or not (by using an old UI memento) depending on their motive. This is an example of a phenomenon described by Acker et al. \cite{acker2020weaponization} where people have used web archives for propagating misinformation. The same can be done for VTR labeled tweets by using mementos before Twitter added this type of label to its old UI.

Archiving Twitter has always had its challenges, and the change in Twitter’s UI has introduced further anomalies \cite{NewTwitterUI:HimarshaKritika} \cite{TwitterRewritesYourURLS:HimarshaKritika}. Sometimes, an archived English language Twitter page will replay in IA either in a different language or with multilingual content, displaying a web page that never existed on the live web. This has been identified as a \emph{cookie violation} \cite{CookiesViolations:Sawood} \cite{alam2019impact}. In Twitter’s old UI mementos, the session cookies present at the time of capturing the Twitter pages impact the language displayed in the resulting mementos. However, we have seen that 
the use of cookies for language is not supported by the new Twitter UI. Therefore, this phenomenon does not impact the new UI mementos as it does for the old UI. They also discovered that Kannada (a regional Indian language) is the 2nd most prominent language in the web archives. We verified this by studying the language distribution on our collected memento dataset and received similar results. 

In IA, the archival datetime (Memento-Datetime) of a web page can be determined from the URI and the banner, but the archival datetime of different sections of a complex web page (composite memento) may or may not be temporally aligned. When archived embedded resources are incorrectly combined with an archived root HTML page to replay a web page that never existed on the live web, the resultant memento is called \emph{temporally violative} \cite{ainsworth2015only}. The tweet section of Twitter's old UI is temporally coherent since the most recent 20 tweets are embedded in the HTML. However, there is a polling request to load new tweets after every 30 seconds and this request populates the Twitter timeline with a new JSON response containing tweets. In old UI memento, this JSON can be from a different datetime making this old UI memento temporarily violative. For Twitter’s new UI, since all the tweets are loaded by an asynchronous JSON response, the new UI mementos are more susceptible to temporal violation.

We focused our study on the change in Twitter's UI that made it difficult to archive Twitter. Similar to Twitter, most websites frequently change their inner structure and some of these changes are not friendly to web archives. For example, Instagram has undergone a similar UI change, but as it was already difficult to archive, the impact was not as noticed. CNN was also difficult to replay for a period of three years due to changes they made in their JavaScript \cite{CNNunarchivable:JohnBerlin}.
Achieving high fidelity in archiving the web and its replay is a difficult task that poses an ever-increasing number of challenges as the web evolves and new practices emerge~\cite{Archiveweb:iilya, alam2019bundle, webrecorder:iilya, Scalecollection:ArchiveIt, Pywb2:iilya}.

\section{Methodology}
For our analysis, we quantified the presence of @realDonaldTrump in web archives between May 1, 2020 and January 8, 2021. We chose these endpoints because in May 2020 Twitter introduced new labels to curb the spread of potentially harmful and misleading content \cite{LabelIntroduction:Twitter}, and on January 8, 2021, Twitter permanently suspended the @realDonaldTrump account.

\subsection{Dataset}
\label{sec:Dataset}
We collected @realDonaldTrump’s account page mementos and tweet page mementos from multiple web archives using  MemGator \cite{alam2016memgator}, a Memento aggregator. To collect the account page mementos, we used @realDonaldTrump's Twitter profile page URL along with its 47 language variations \cite{CookiesAreWhy:SawoodPlinio} \cite{LinksMystery:dshr}. We collected  total of 64,719  account page mementos between May 1, 2020 and January 8, 2021. 

To collect tweet page mementos, we first acquired the URI-Rs of tweets from @realDonaldTrump in our selected time frame. We used three methods to ensure that we were collecting the maximum possible archived tweets. 
\begin{enumerate}
\item For our first method, we used the dataset  collected by Summers \cite{TrumpsTweets:Ed} by issuing a query to the Internet Archive's CDX API\footnote{https://github.com/internetarchive/wayback/tree/master/wayback-cdx-server/} for the URL prefix https://twitter.com/realDonaldTrump/status/\{tweet-id\}, a method first introduced by Siddique \cite{SearchingWebArchives:MNS}. This dataset consisted of 16,043,553 mementos of Trump’s tweets and 57,292 unique tweets. From these 57,292 unique tweets, we extracted 8,708 tweets that were from our desired time frame. We also collected different URL forms for each of those tweets using the  CDX API. We verified the creation dates of these unique tweets using the TweetedAt tool \cite{TweetedAt:NaumanSawood} to make sure that they were tweeted between May 1, 2020 and January 8, 2021. 

\item For our second method, we scraped our assembled account page mementos. We collected 7,747 tweets: 4,613 tweets authored by @realDonaldTrump and 3,134 retweets by @realDonaldTrump authored by other users. 

\item In the third method, we used 3,569 archived tweet JSONs to extract 3,194 tweets. We requested the URI-T of the tweet JSON of @realDonaldTrump's account page\footnote{http://bit.ly/tweetJSONtimemap} from the Wayback Machine. We downloaded these mementos of tweet JSON, each containing 19 tweets. 
\end{enumerate}

We compared and combined the tweets gathered by each method, resulting in 8,709 unique tweets (i.e., we only found one new tweet from the second and third methods) using MemGator. We found 190,194 different URL forms for each tweet. In total, we collected 1,554,407 mementos for each tweet URL variation. Table~\ref{tab:tab_1} shows the distribution of those account page and tweet page mementos across seven different web archives.

\begin{table}[!h]
\large
  \caption{Distribution of 1,554,407 account page and tweet page mementos across 7 different archives}
\setlength\tabcolsep{0.5pt}
\setlength\extrarowheight{3.5pt}
\begin{tabularx}{0.5\textwidth}{*{5}{Z}}
\toprule
\thead{Web Archives}  & 
\thead{Account Page \\ Mementos} & 
\thead{Tweet Page \\ Mementos}\\
\midrule
archive.today             & 2,622                 & 5,887               \\
\midrule
perma.cc               & 10                    & 140                 \\
\midrule
swap.stanford.edu      & 2                     & 308                 \\
\midrule
archive-it.org & 2,498                 & 7,821               \\
\midrule
vefsafn.is     & 1,371                 & 99                  \\
\midrule
webarchive.org.uk  & 2                     & 38                  \\
\midrule
web.archive.org        & 58,214                & 1,282,734           \\
\midrule
Total Mementos         & 64,719                & 1,297,027\\          
 \bottomrule
\end{tabularx}
  \label{tab:tab_1}
\end{table}

\subsection{Old UI vs. New UI }
We used the content-length from the HTTP response of each memento to distinguish between the old Twitter UI and the new Twitter UI, as the content-length of an old UI memento is significantly larger than that of a new UI memento \cite{TwitterWasAlreadyDifficult:KritikaHimarsha}. As discussed in Section \ref{sec:background}, the new UI relies largely on API calls to fill out the content of the page, whereas the content in the old UI was embedded in the base HTML page. We used two methods for collecting the content-length of the acquired URI-Ms. For Wayback Machine URI-Ms, we used IA’s CDX API to collect the recorded content-length of the mementos. We used this method for our 1,340,948 Wayback URI-Ms, as sending an HTTP request for each of these mementos would take much longer and tax the archive. 
For the significantly fewer URI-Ms collected from other web archives, we made an HTTP request to each URI-M to collect their respective content-length. 

Along with content-length, we also noted  the HTTP response status code of each memento, excluding the 37,383 URI-Ms that received an HTTP status code other than ‘200 OK’.  We then separated the collected 64,364 account page mementos into 3,615 new UI mementos and 60,749 old UI mementos based on their content-length. Similarly, we successfully separated the 1,259,999 tweet page mementos into 1,175,217 old Twitter UI and 84,782 new Twitter UI. Table~\ref{tab:tab_2} shows the distribution of mementos with ‘200 OK’, across six different web archives. We excluded the mementos from webarchive.org.uk that returned the HTTP ‘451 Unavailable For Legal Reasons’ response code. 

\begin{table*}[!h]
\large
\centering
  \caption{Distribution of account page and tweet page mementos across 6 different archives}
\setlength\tabcolsep{0.7pt}
\setlength\extrarowheight{4pt}
\begin{tabularx}{0.9\textwidth}{*{7}{Z}}
\toprule
\thead{Web \\ Archives}  & 
\multicolumn{3}{c}{\thead{Account Page \\ Mementos}} & 
\multicolumn{3}{c}{\thead{Tweet Page \\ Mementos}} \\
& \thead{Total} & \thead{New UI} & \thead{Old UI} &
\thead{Total} & \thead{New UI} & \thead{Old UI}\\ 
\midrule
archive.today & 2,621 & 33 & 2,588 & 5,887 & 5,632 & 255 \\
\midrule
perma.cc & 10 & 9 & 1 & 140 & 133 & 7 \\
\midrule
swap.stanford\\.edu & 2 & 0 & 2 & 308 & 0 & 308 \\
\midrule
archive-it.org & 2,498 & 17 & 2,481 & 7,696 & 475 & 7,221 \\
\midrule
vefsafn.is & 1,371 & 1,152 & 219 & 99 & 98 & 1 \\
\midrule
web.archive\\.org & 57,862 & 2,404 & 55,458 & 1,245,869 & 78,397 & 1,167,472\\          
 \bottomrule
\end{tabularx}
  \label{tab:tab_2}
\end{table*}

We separated the number of old UI and new UI mementos for account pages in each month from May 2020 to January 2021. Figure~\ref{fig:UIdistribution_barplot}(a) shows that there are many more  old UI mementos than new UI mementos each month. We observed a similar pattern for the tweet page mementos (Figure~\ref{fig:UIdistribution_barplot}(b)). After Twitter stopped supporting the old UI in June 2020, we expected to see more mementos of the new UI. However, we observed otherwise since web archives continued to archive the old UI due to problems in archiving the new UI.

\begin{figure}[htbp]
\centerline{\includegraphics[width=0.5\textwidth]{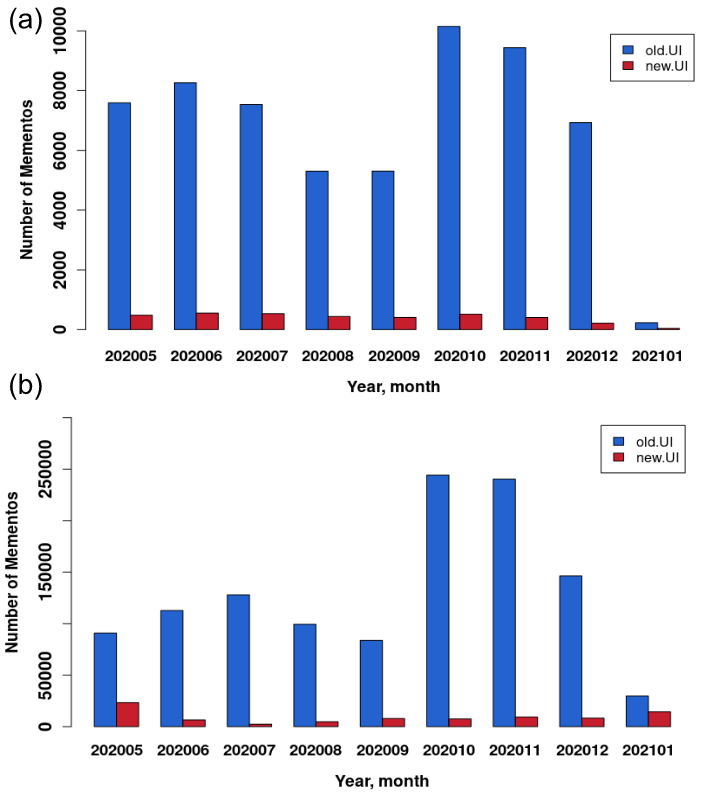}}
\caption{Distribution of old UI and new UI for account page mementos (a) and tweet page mementos (b) across each month from May 2020 to  January 2021}
    \label{fig:UIdistribution_barplot}%
\end{figure}

We calculated the number of old UI mementos and new UI mementos for each of the 8,709 tweets present in our dataset. We observed that 
9.46\% (824) of tweets have only old UI mementos, 45.79\% (3,988) have only new UI mementos, and 44.54\% (3,897)  have both old and new UI mementos. This shows how for some tweet IDs, web archive users would have to depend only on new UI mementos, which are problematic. 

\subsection{Temporal Violations}
We quantified the temporal spread of each of the five sections in the new UI: Tweet feed, Bio, Media timeline, ``You might like”, and ``What's happening” (Figure~\ref{fig:temporal_range}). Temporal spread\cite{ainsworth2015only} is the difference between the earliest and latest Memento-Datetimes of the JSON responses populating the sections. For each of the new UI account page mementos in our dataset, we obtained the mementos of the JSON responses for each of the five sections. We noticed that only 1,790 mementos out of 3,615 new UI account page mementos had archived JSON responses for all five sections. We refer to these mementos as \emph{complete new UI mementos} and the rest of the 1,825 mementos as failed new UI mementos. 

\begin{figure*}[htbp]
\centerline{\includegraphics[width=1\textwidth]{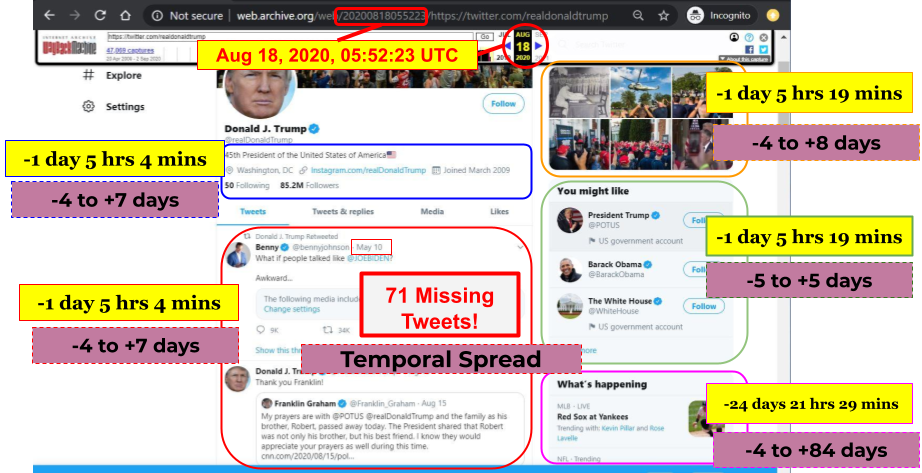}}
\caption{Twitter new UI memento (archived on 2020-08-18T05:52:23Z) has 71 missing tweets. The archived JSON for tweets, bio, media, “You might like”, and “What’s happening” have memento datetimes from the past. The range (min and max) of time delta for each section of a Twitter page is also shown. \url{https://web.archive.org/web/20200818055223/https://twitter.com/realdonaldtrump}}
\label{fig:temporal_range}
\end{figure*}

Although our dataset comprises new UI mementos from five different archives, all of the complete new UI mementos are from the Wayback Machine. In the Wayback Machine, if a resource is unavailable at the requested time, the request is redirected to the same URI-R nearest to the requested Memento-Datetime. This property ensures that replay of the new UI is complete, but it can also create a replay of a new UI memento that is temporally violative if this spread is too large.

To quantify the temporal spread in our 1,790 complete new UI mementos, we calculated the time delta, which is the difference between the requested Memento-Datetime of the root HTML and the Memento-Datetime of the JSON response populated during the replay. We separated the time delta for the mementos with tweet JSON populating from the future (Table~\ref{tab:tab_4}) and the past (Table~\ref{tab:tab_5}) along with their min, max, mean, median, and standard deviation in seconds and hours. 

\begin{table*}[!h]
\fontsize{10}{7.2}
\centering
  \caption{The min, max, mean, median, and standard deviation of the time delta for the memento with JSON responses populating from the \emph{future}. The table shows the breakdown for each section in Twitter's new UI.}
\setlength\tabcolsep{0.4pt}
\setlength\extrarowheight{5pt}
\begin{tabularx}{0.9\textwidth}{*{7}{Z}}
\toprule
\thead{Section}  & 
\thead{Memento\\ Count} & 
\thead{Min \\in sec} & 
\thead{Max \\in sec\\ (hr)} & 
\thead{Median \\in sec\\ (hr)} & 
\thead{Mean \\in sec\\ (hr)} & 
\thead{Sd \\in sec\\ (hr)}\\
\midrule
Tweet & 957 & 1 & 622752 (173.0) & 6627 (1.8) &   44528 (12.4) & 95190 (26.4) \\
\midrule
Bio & 977 & 1 & 539158 (149.8) & 7925 (2.2) & 44948 (12.5) & 92633 (25.7) \\
\midrule
Media & 953 & 1 & 622754 (173.0) & 7119 (2.0) & 45553 (12.7) & 95428 (26.5) \\
\midrule
You might like & 948 & 1 & 368455 (102.3) & 7370.5 (2.0) & 33420 (9.3) & 55020 (15.3) \\
\midrule
What's happening & 926 & 1 & 7203844 (2001.1) & 4422.5 (1.2) & 44030 (12.2) & 249122 (69.2)\\          
 \bottomrule
\end{tabularx}
  \label{tab:tab_4}
\end{table*}

\begin{table*}[!h]
\fontsize{10}{7.2}
\Centering
  \caption{The min, max, mean, median, and standard deviation of the time delta for the memento with JSON responses populating from the \emph{past}. The table shows the breakdown for each section in Twitter's new UI.}
\setlength\tabcolsep{0.45pt}
\setlength\extrarowheight{5pt}
\begin{tabularx}{0.9\textwidth}{*{7}{Z}}
\toprule
\thead{Section}  & 
\thead{Memento\\ Count} & 
\thead{Min \\in sec} & 
\thead{Max \\in sec\\ (hr)} & 
\thead{Median \\in sec\\ (hr)} & 
\thead{Mean \\in sec\\ (hr)} & 
\thead{Sd \\in sec\\ (hr)}\\
\midrule
Tweet & 833 & 4 & 348431 (96.8) & 23593 (6.6) & 51221 (14.2) & 66089 (18.4) \\
\midrule
Bio & 809 & 5 & 348431 (96.8) & 21855 (6.1) & 48060 (13.4) & 63757 (17.7) \\
\midrule
Media & 837 & 4 & 348430 (96.8) & 25384 (7.1) & 52318 (14.5) & 66026 (18.3) \\
\midrule
You might like & 842 & 3 & 365722 (101.6) & 24908 (6.9) & 56131 (15.6) & 74256 (20.6) \\
\midrule
What's happening & 860 & 2 & 348430 (96.8) & 15032.5 (4.2) & 45442 (12.6) & 63576 (17.7)\\          
 \bottomrule
\end{tabularx}
  \label{tab:tab_5}
\end{table*}

Figure~\ref{fig:ecdf_timedelta} illustrates the distribution of the time delta for each of the five UI sections over all of the complete new UI pages. The x-axis represents the time delta, and the y-axis represents the cumulative probability of time delta. For each of the five sections, around 70\% of the time delta values are between $-12$ hours to $+12$ hours. For the ``What's happening” section, there is one exception where the JSON response is from 83 days in the future from the root memento.

\begin{figure*}[htp]
    \centering
    \subfloat[\centering]{{\includegraphics[width=6cm]{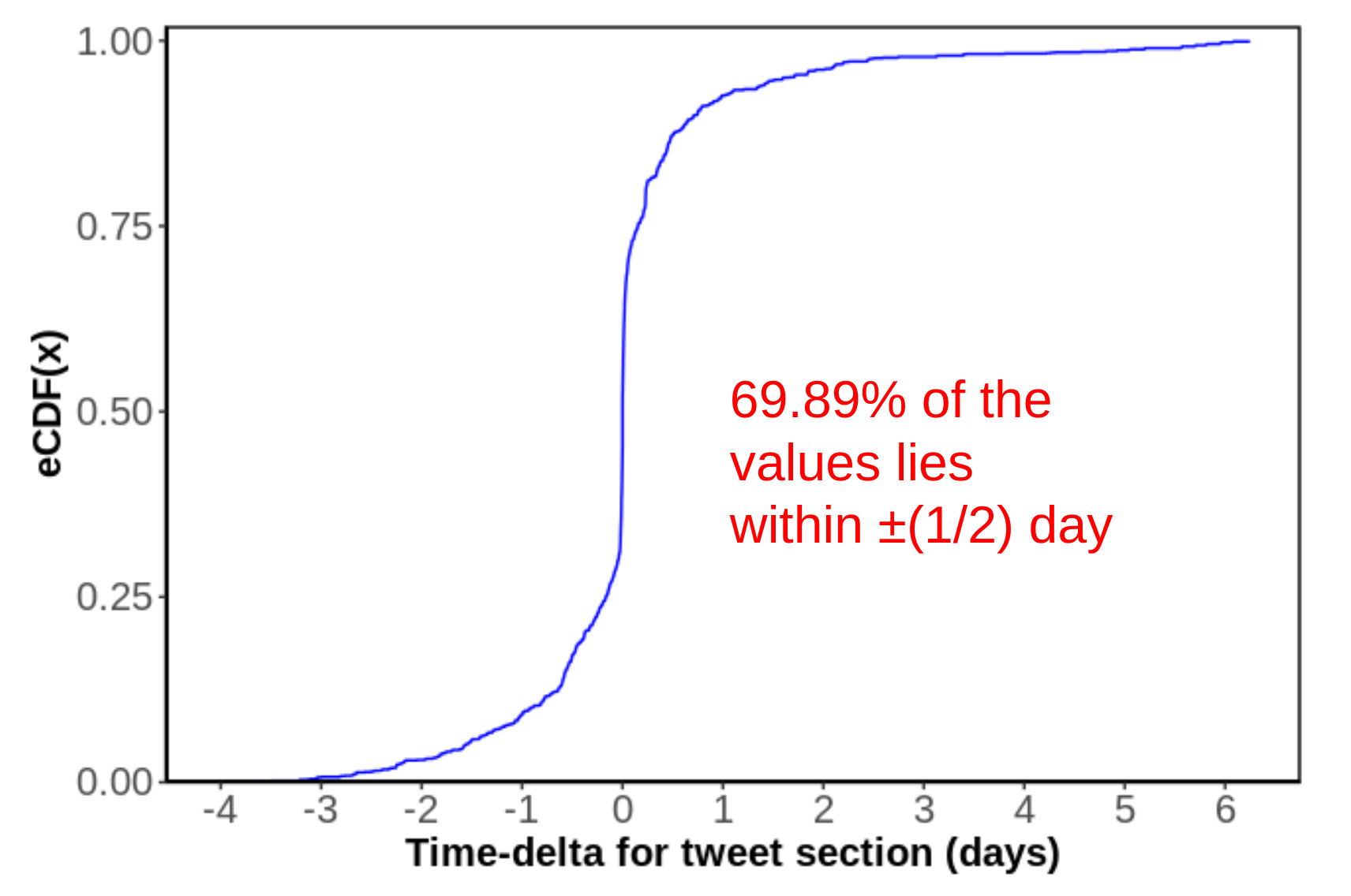} }}
    \subfloat[\centering]{{\includegraphics[width=6cm]{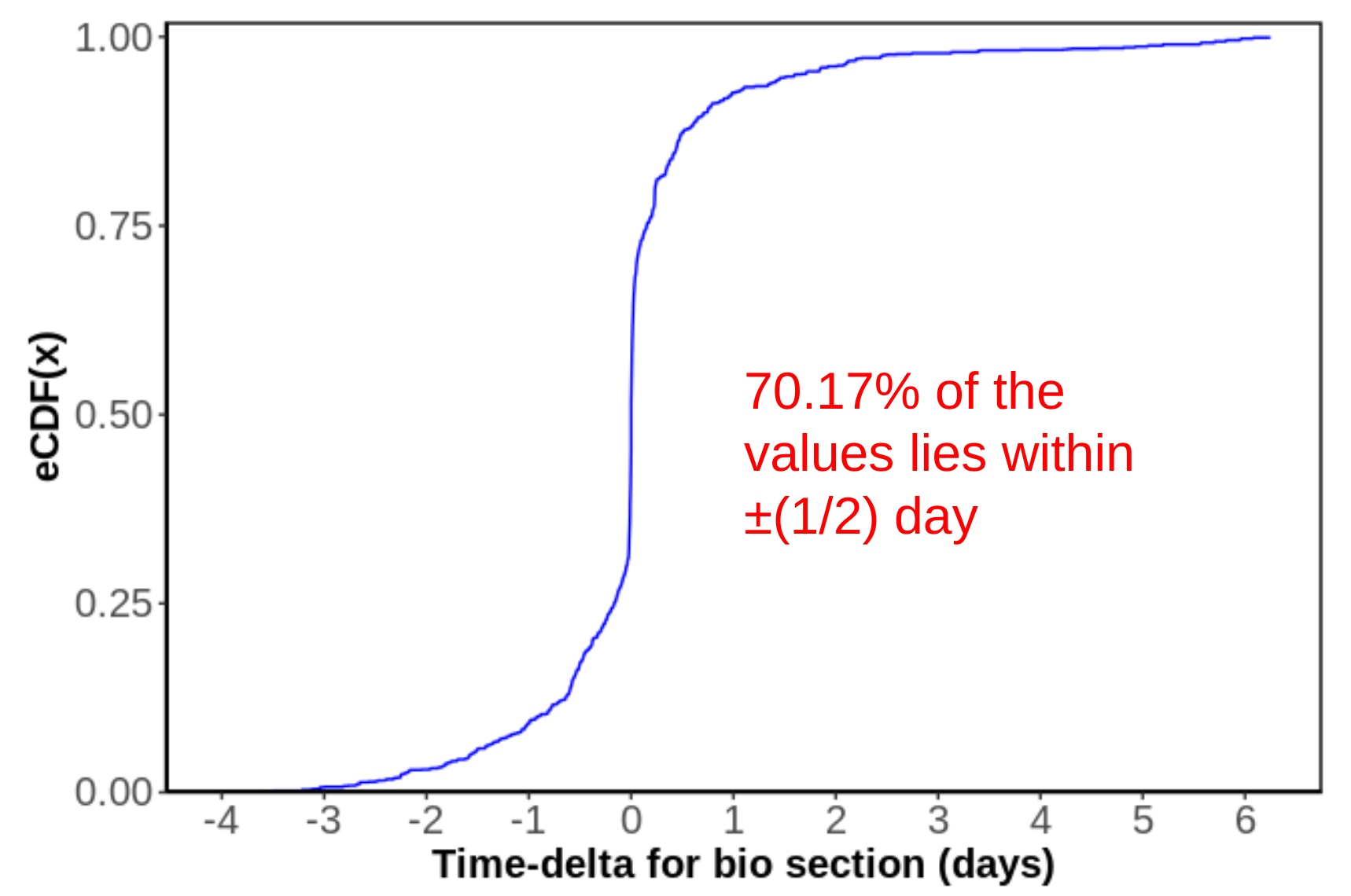} }}%
 \subfloat[\centering]{{\includegraphics[width=6cm]{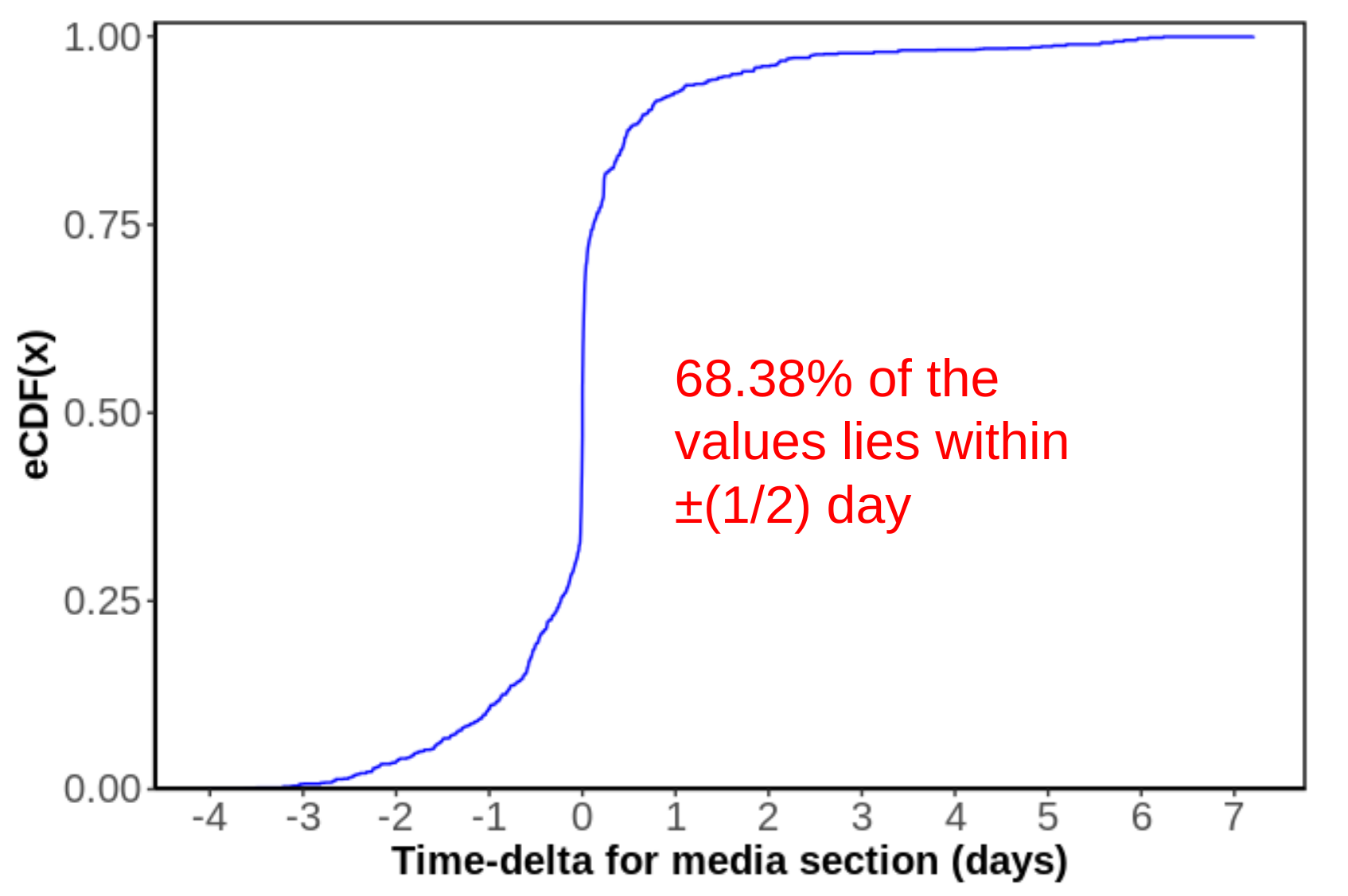} }}
    \newline
    \subfloat[\centering]{{\includegraphics[width=6cm]{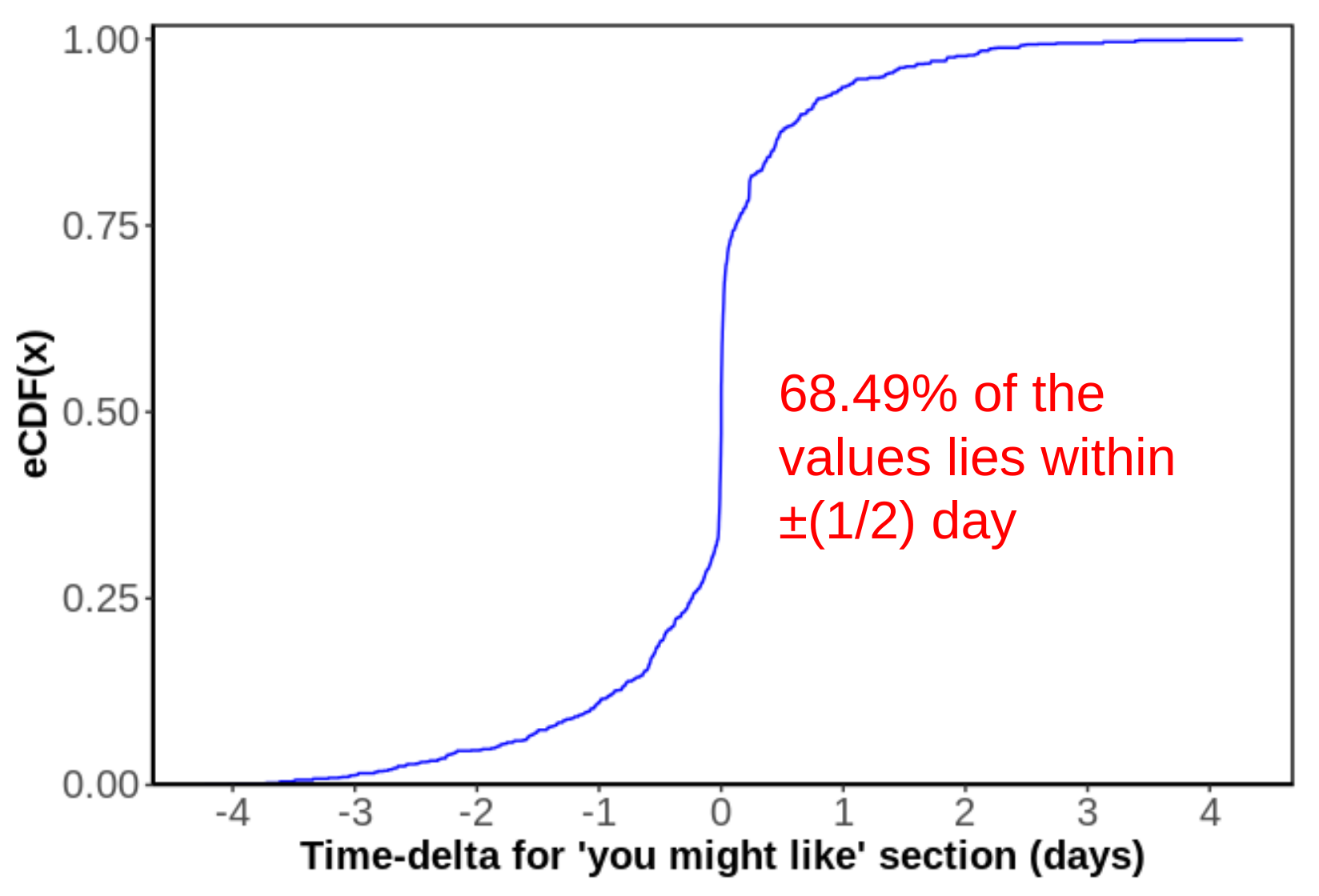} }}%
 \subfloat[\centering]{{\includegraphics[width=6cm]{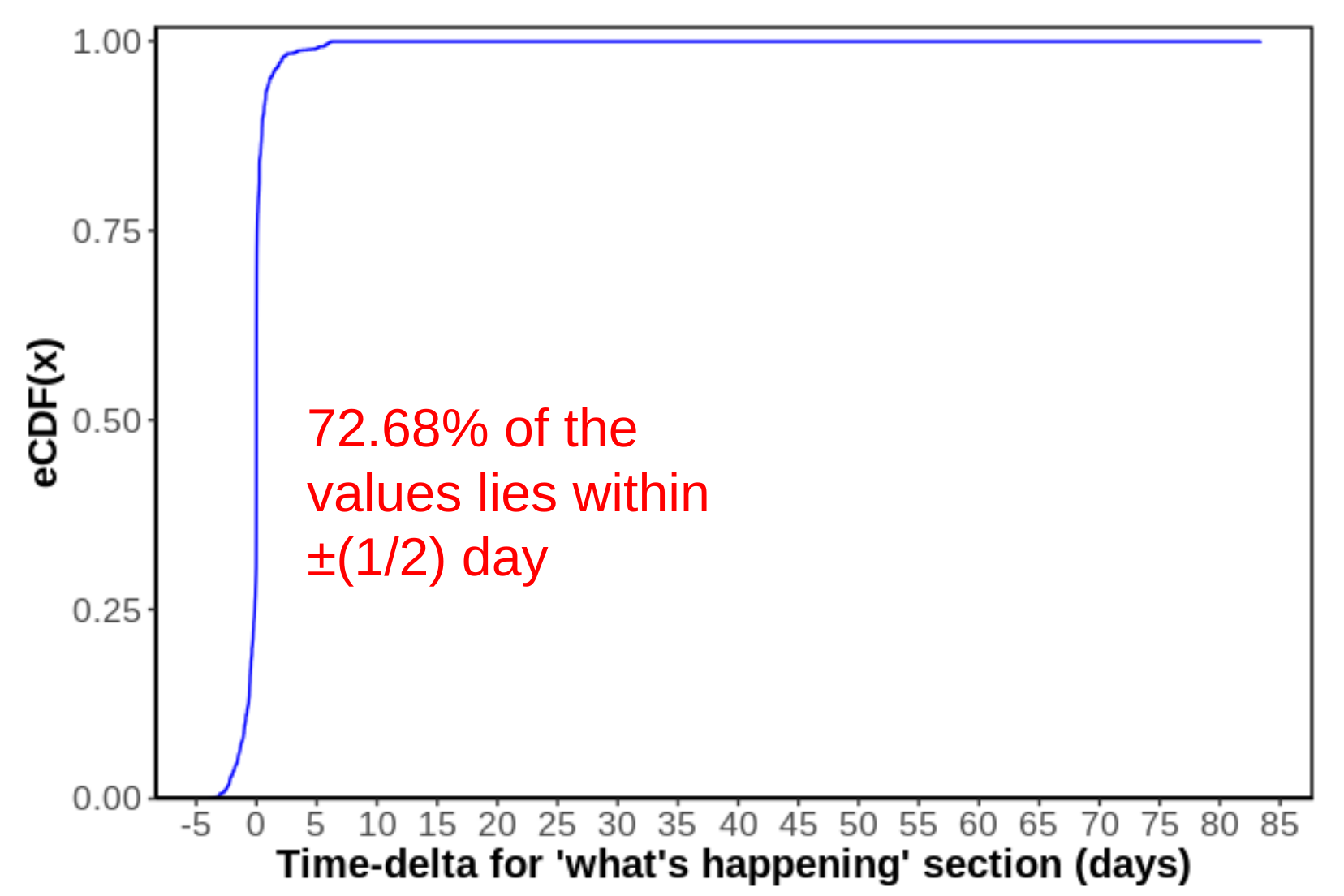} }}
    \caption{Empirical Cumulative Distribution Function (ECDF) graph for each section in Twitter's new UI. The x-axis denotes the time-delta for (a) Tweet, (b) Bio, (c) Media, (d) You might like, and (e) What’s happening sections and the y-axis denotes the cumulative probability. }%
    \label{fig:ecdf_timedelta}%
\end{figure*}

Even though each section of the Twitter page has a wide range (min and max) of time delta, there are different implications. For example, the spread in the ``Tweet'' section is more impactful than the spread in the ``Bio'' section because most people do not frequently edit their bio information. Similarly, unlike the ``Media"  and ``You might like” sections, the ``What's happening" section changes frequently and reflects current events. 
We also empirically determined that it takes approximately 5-6 seconds for the new UI page to completely load on the live web under normal conditions; a larger time delta would not reflect a likely live web experience.

\subsection{Labels}
\label{sec:labels}
We examined the tweets of @realDonaldTrump that were labeled by Twitter, both ``Fact-check” and VTR labels. We also looked at the presence of these labeled tweets in IA. According to the timeline illustrated in Figure~\ref{fig:labeltimeline}, we would expect a memento with datetime after $T_2$ to have a label on it. Labels are a feature of the new UI, so archives that were capturing the old UI do not have the labels. However, we verified that Twitter added the VTR label to its old UI by at least August 26, 2020, but the ``Fact-check” label has not been added to the old Twitter UI as of June 2021. This means that the ``Fact-check” label is missing from the mementos of labeled tweets even after $T_2$, when it is captured by public web archives archiving old UI. 

We collected all tweets from @realDonaldTrump that were labeled by Twitter using three datasets:
\begin{itemize}
\item thetrumparchive.com: A public dataset that was compiled by checking Twitter every 60s to record all of @realDonaldTrump’s tweets that were authored, retweeted, or quote tweeted by him. We used a subset of tweet IDs from this dataset by filtering the tweet IDs that were marked as labeled in its database. 

\item factba.se: A dataset of Trump's labeled tweets. We used Selenium\cite{selenium}
to scroll down the factba.se  page and then downloaded the HTML. We parsed this HTML file to extract the tweet ID and the type of tweet (Tweet or RT). 

\item twitterlabels6: A dataset\cite{twitterlabels_2021} of 6 months of labeled tweets that we collected and used in our previous work \cite{TwitterAddedLabels:HimarshaKritika}. This contains data between May 26, 2020 and November 26, 2020 and includes the date, tweet ID, whether or not it is a retweet, and the type of label according to our categorization.
\end{itemize}

\begin{table}[!h]
\fontsize{8}{7.2}
\caption{Total number of labeled tweets, type of tweet, and the date range in each of the three datasets.}
\setlength\tabcolsep{0.5pt}
\setlength\extrarowheight{3.5pt}
\begin{tabularx}{0.5\textwidth}{*{6}{Z}}
\toprule
\thead{Dataset}  & 
\thead{Number of \\ labeled \\ tweet IDs } & 
\thead{Tweets} & 
\thead{Retweets} & 
\thead{First \\labeled \\tweet} & 
\thead{Last \\labeled\\ tweet} \\
\midrule
thetrump\\archive.com & 303 & 249 & 54 & 2020-05-26 & 2020-12-10 \\
\midrule
factba.se & 471 & 385  & 86  & 2020-05-26 & 2021-01-06\\
\midrule
twitterlabels6 & 167 & 160  & 7 & 2020-05-26 & 2020-11-26 \\
 \bottomrule
\end{tabularx}
    \label{tab:tab_7}
\end{table}

Table~\ref{tab:tab_7} shows a summary of the data collected from each of the three datasets. We gathered a total of 476 unique tweet IDs from the three datasets. All of these tweet IDs are present in the main thetrumparchive.com dataset although not all are marked as labeled there. 
\begin{figure*}[htbp]
\centerline{\includegraphics[width=0.65\textwidth]{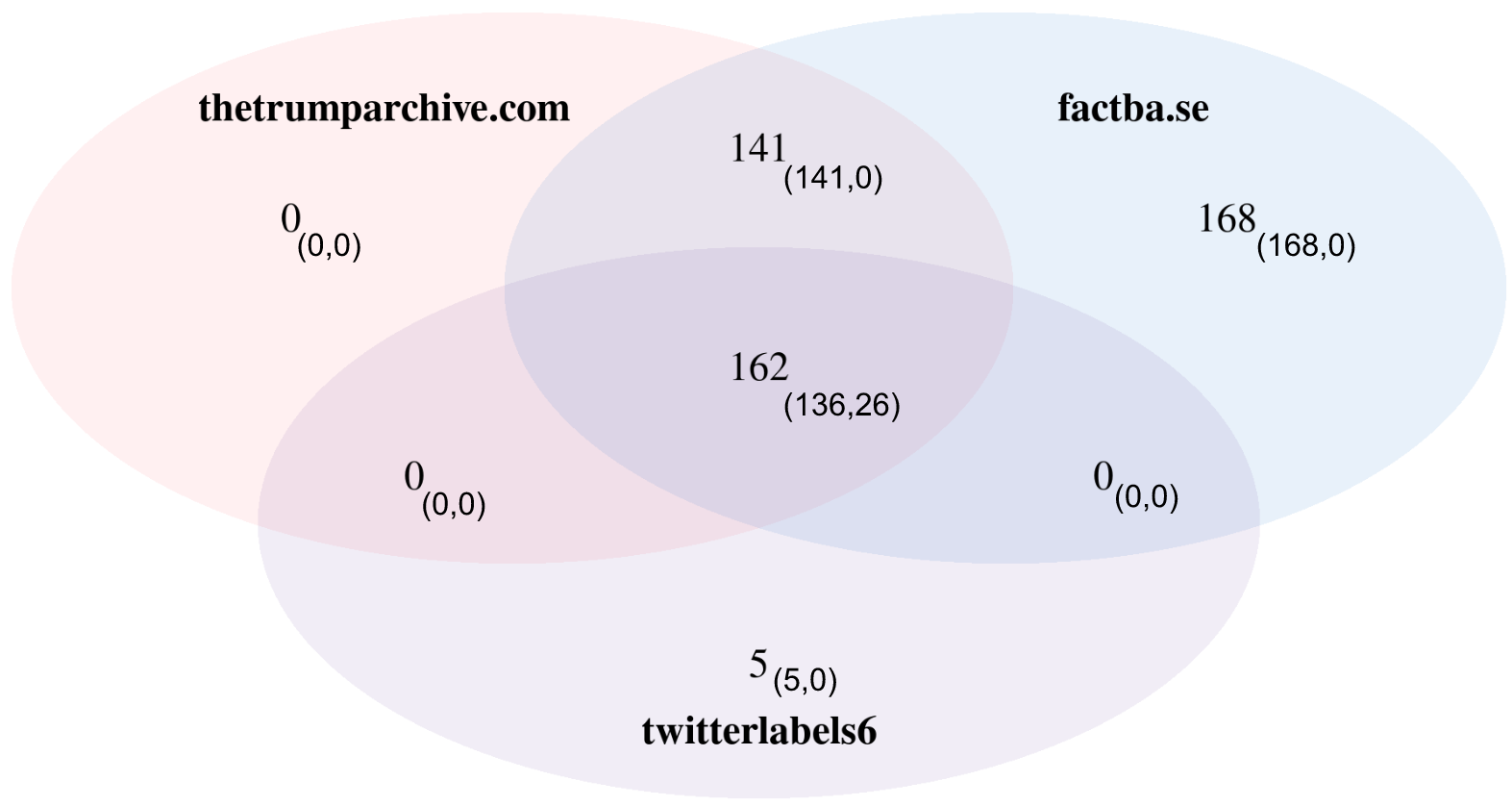}}
\caption{Venn diagram illustrating the relationship between the datasets of labeled tweets from thetrumparchive.com, factba.se, and our dataset. It reports 476 tweet IDs in the form X(a,b), where X is total, a is ``Fact-check” labeled, b is VTR labeled}
\label{fig:label_venndiagram}
\end{figure*}

 Figure~\ref{fig:label_venndiagram} shows the relationship between the labeled tweet IDs present in each of the three datasets. The factba.se database has all of the tweet IDs from thetrumparchive.com and has all but five tweet IDs from our dataset. Those five tweets from our dataset were not present in thetrumparchive.com either. This shows that none of the databases have a complete set of labeled tweet IDs. There may be more labeled tweets from the  @realDonaldTrump account that were not observed by any of the three datasets. However, in our analysis, we considered these 476 tweet IDs as the complete set of labeled tweets from @realDonaldTrump.

We categorized each of these tweets as having either a ``Fact-check” label or a VTR label. There were 450 ``Fact-check” labeled tweet IDs and 26 VTR labeled tweet IDs. Figure~\ref{fig:label_venndiagram} also illustrates the (Fact-check, VTR) labeled tweets distribution. We can see that the VTR labeled tweets were present in all three datasets. 

\subsubsection{Fact-check Label}
We collected all the old UI and new UI mementos of individual tweet pages in IA for each of the 450 tweets with a ``Fact-check" label. 
Out of the 122,891 mementos collected,  95.12\% (116,897) were old UI mementos and only 4.88\% (5,994) were new UI mementos. Since ``Fact-check" labels are not a part of Twitter’s old UI, we only consider the new UI mementos. We evaluated the status of these new UI mementos as well as the presence of labels in Section \ref{sec:evaluation}. There is no information available  on the time difference between $T_1$ and $T_2$ for these tweets, which makes it impossible  to determine if the memento was captured at $T_{3-1}$ or $T_{3-2}$.  If we assume that these tweets were already labeled on the live web at the time of capture by IA (captured after $T_2$), all the old UI mementos (95.12\% of the total) have missing labels, because this type of label was not added to Twitter’s old UI. Otherwise, this number will vary depending on when the memento was created (at $T_{3-1}$ or $T_{3-2}$), since only mementos captured after $T_2$ would have had labels. 

\subsubsection{Violated Twitter Rules (VTR) Labels}
As opposed to ``Fact-check” labels, these VTR labeled tweets had their engagement limited to only quote tweeting after $T_2$. We used this property to identify the tweets by @realDonaldTrump that had zero retweet count and favorite count as VTR labeled tweets. We were able to collect 26 such labeled tweets. Twitter did not support this kind of label on the old UI until August 26, 2020. For these 26 labeled tweets, we were able to obtain 74,883 tweet page mementos, where 95.12\% (74,324) are old UI mementos and 4.88\% (559) are new UI mementos. 

Table~\ref{tab:tab_8} shows a summary of all the tweet IDs belonging to the @realDonaldTrump account that were labeled by Twitter as well as the presence of those tweet pages in web archives.
\begin{table}[!h]
\large
\caption{Number of mementos for tweets by @realDonaldTrump that were labeled in IA.}
\setlength\tabcolsep{0.5pt}
\setlength\extrarowheight{3.5pt}
\begin{tabularx}{0.5\textwidth}{*{4}{Z}}
\toprule
\thead{Type of \\ Label}  & 
\thead{Tweets} & 
\thead{Old UI \\Mementos} & 
\thead{New UI \\Mementos} \\ 
\midrule
Fact-check & 450 & 116,897 & 5,994 \\
\midrule
VTR & 26 & 74324 & 559 \\
\midrule
Total & 476 & 191,221 & 6,553 \\
 \bottomrule
\end{tabularx}
\label{tab:tab_8}
\end{table}
Out of the total number of mementos for both types of labels (197,774), 96.69\% (191221) are of Twitter’s old UI and only 3.31\% are new UI mementos. This shows how IA is still primarily capturing Twitter's old UI. 

\section{Evaluation}
\label{sec:evaluation}
\subsection{Comparison between web archives and thetrumparchive.com}
In our study, we considered all @realDonaldTrump tweets from the time of Twitter’s new UI until account suspension. We also compared the tweet IDs that we obtained from the web archives using the three methods mentioned in Section \ref{sec:Dataset} against the thetrumparchive.com dataset. We extracted 8,719 tweets created between May 1, 2020 and January 8, 2021 from thetrumparchive.com. Of these, 28 tweet IDs were unique to thetrumparchive.com and 18 tweet IDs unique to IA. To verify these two different sets of IDs, we ran both MemGator and CDX searches for these IDs. Surprisingly, we found at least one memento in IA for all 28 IDs that we thought were unique to thetrumparchive.com. Because CDX responses may vary each time a request is made, there is the possibility that those mementos were not in the CDX index when we complied our dataset. Of those 28 tweet IDs, 
\begin{itemize}
\item 18 tweet IDs have mementos with a response code of 301, 302, or 404
\item 10 tweet IDs have at least one memento with a response code of 200.
\end{itemize}

Of the 18 IDs that are present in IA but not in thetrumparchive.com, 
\begin{itemize}
\item 16 of those have at least one memento resulting in a response code of 200.
\item 2 of those have mementos that were only resulting in a response code of 301. We identified these two IDs to be original tweet IDs that belong to a different account that @realDonaldTrump has retweeted.
\end{itemize}

We have verified each of the 18 tweet IDs that were unique to web archives, and Table~\ref{tab:tab_9} shows how we have categorized each tweet ID according to its type.

\begin{table}[!h]
\fontsize{11}{7.2}
\caption{Summary of the 18 tweet IDs that are present in IA but not in thetrumparchive.com}
\setlength\tabcolsep{0.5pt}
\setlength\extrarowheight{3.5pt}
\begin{tabularx}{0.5\textwidth}{*{3}{Z}}
\toprule
\thead{S/N} & 
\thead{Tweet ID} & 
\thead{Category} \\
\midrule
1 & 1258113511730884611 & {Original tweet ID}\\
\midrule
2 & 1267997425463111680 & {Original tweet ID}\\
\midrule
3 & 1268356663259660293 & {Retweet ID}\\
\midrule
4 & 1280863277543800832 & {Retweet ID}\\
\midrule
5 & 1281192788588183558 & {Retweet ID}\\
\midrule
6 & 1334001254012497923 & {Retweet ID}\\
\midrule
7 & 1270410940132069385 & {Tweet ID}\\
\midrule
8 & 1280130505829224448 & {Tweet ID}\\
\midrule
9 & 1322863297025220608 & {Tweet ID}\\
\midrule
10 & 1325173848396898305 & {Tweet ID}\\
\midrule
11 & 1337526501017858055 & {Tweet ID}\\
\midrule
12 & 1338329628616830977 & {Tweet ID}\\
\midrule
13 & 1281071985712390145 & Apocryphal\\
\midrule
14 & 1281634152195018757 & Apocryphal\\
\midrule
15 & 1311785763559797506 & Apocryphal\\
\midrule
16 & 1311892190680014848 & Apocryphal\\
\midrule
17 & 1313065153725627347 & Apocryphal\\
\midrule
18 & 1324855496722026493 & Apocryphal\\
 \bottomrule
\end{tabularx}
    \label{tab:tab_9}
\end{table}

In table~\ref{tab:tab_9}, the tweet IDs 1 and 2 are not from @realDonaldTrump account. However, through web archives and other third-party services like didtrumptweetit.com, twitterdonald.wordpress.com, allthepresidentswords.wordpress.com, and facebook.com/SoonerPolitics/, we could verify that four of the tweet IDs  were retweet IDs and 6 were tweet IDs from the @realDonaldTrump account, which should have been included in thetrumparchive.com. There were another six tweet IDs that may be valid tweet IDs, but there is no evidence in other third-party datasets that we could use for verification. Although mementos for these tweets are present in web archives, unfortunately, they are all non-working new UI mementos. Hence, we are unable to use those mementos to verify these tweet IDs by looking at the content. Because of the doubtful authenticity of these tweet IDs, we categorized them as \emph{apocryphal}. 

Even with a large number of public third-party databases that could be used for verification for the @realDonaldTrump account,  we can see there are still a few apocryphal tweet IDs. In the case of other accounts that were suspended along with Trump, there are little or no ground truth datasets to compare against, so web archives will be the primary source of information, at least for relatively popular accounts.

\subsection{Temporal Violations}
We analyzed the extent of temporal violations in our 1790 complete new UI account page mementos. We used the time delta between the memento datetimes of the root HTML and the JSON populating the tweet section to analyze the number of missing tweets in the memento. We found that out of 1,790 mementos, 48.8\% were temporally violative, i.e. they had at least one missing tweet during the time delta.
We first analyzed the new UI mementos that are vulnerable to temporal violations from the past, meaning that the tweet JSON had an earlier memento datetime than the root HTML. We counted the number of tweets that occurred during the time delta using thetrumparchive.com and classified these as missing tweets on the account page memento. Out of 833 mementos whose tweet JSON are populated from the past, 67\% were temporally violative.
We then analyzed the temporal violations in mementos with tweet JSON populated from the future. To measure the number of tweets that the memento is off by, we obtained the tweets that were created during the time delta using thetrumparchive.com and counted the tweets that should not have been present in the tweet JSON.  Out of 958 mementos whose tweet JSON are populated from the future, 66.6\%  were temporally violative.   

Figure~\ref{fig:timedeltaVmissingtweets} shows the correlation between the number of tweets each memento is off by and the memento's time delta. The graph shows that there is a linear relationship, showing that as the time delta increases, the number of tweets the memento is off by also increases. However, this relationship may only hold for highly active accounts such as @realDonaldTrump. This can be seen by observing the outliers in the graph. For example, one outlier shows that 115 tweets are missing in under 7.7 hours. For accounts with less activity, the time delta would have to be higher before inconsistencies in new UI mementos due to temporal violation would be apparent. 
\begin{figure}[htbp]
\centerline{\includegraphics[width=0.5\textwidth]{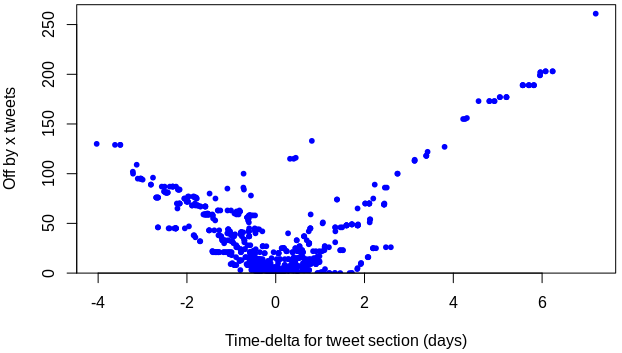}}
\caption{The correlation between the time delta and number of missing (negative delta) or future (positive delta) tweets in each memento
}
\label{fig:timedeltaVmissingtweets}
\end{figure}

We similarly examined the temporal violations in the ``What’s happening'' section of Twitter's new UI account page mementos. This section polls for new updates every five minutes, but this does not mean that all the stories in the section get updated. We quantified the possible number of updates a user may miss in a new UI memento. For example, a user will miss 11 updates if the JSON populating the ``What's happening" section is off by an hour. We found a linear relationship between the missing updates and the time delta. 

\subsection{Labels}
We evaluated the presence of Twitter labels in web archives as compared to the live web. In other sections, we have used web archives to collect the data and then compared it against the ground truth, but it is not possible to locate a complete set of labeled tweet IDs by using only the web archives. This is because a) fact-check labels are not added to the old UI, and b) we are unable to use new UI mementos, since most of them are not working. 

We used the three existing third-party datasets as discussed in Section \ref{sec:labels} to compile the set of labeled tweet IDs from @realDonaldTrump. By using the union of tweet IDs obtained from the three datasets, we obtained the new UI mementos in IA for those tweets. We also identified how many of these are complete mementos as well as checking for the presence of the two types of labels. 

There were 450 ``Fact-check” labeled tweets with  5,994 new UI mementos, and there were 26 ``Violated Twitter Rules” labeled tweet IDs with 559 new UI mementos. We quantified how many of these new UI mementos are replaying properly as well as the presence of labels in the mementos. We extracted the URI-M of the archived tweet JSON from each new UI memento using Selenium Wire\cite{selenium-wire}. This JSON request and its responses helped us determine if the memento is replaying the tweets, and we then checked for the presence of the labels in the tweet JSON. We performed two iterations of this process to extract labels from mementos, and each iteration gave us different results. If a memento was working in at least one iteration, we considered it as a working memento. Similarly with the labels, if we found a label in at least one iteration, we marked the label as present.

Table~\ref{tab:tab_10} shows the summary of mementos for labeled tweet IDs in IA according to their completeness in replaying as well as the presence of labels.
\begin{table}[!h]
  \caption{Number of mementos, their status, and presence of label in the new UI mementos for Trump's tweet IDs}
\setlength\tabcolsep{0.5pt}
\setlength\extrarowheight{4pt}
\begin{tabularx}{0.5\textwidth}{*{5}{Z}}
\toprule
\thead{Type of \\ Label}  & 
\thead{Tweets} & 
\thead{New UI \\Mementos} & 
\thead{Working \\Mementos} & 
\thead{Presence\\ of Label} \\
\midrule
Fact-check & 450 & 5,994 & 1,615 & 967 \\
\midrule
VTR & 26 & 559 & 272 & 213 \\
\bottomrule
\end{tabularx}
\label{tab:tab_10}
\end{table}
Out of all the ``Fact-check” labeled new UI mementos at IA (5,994), at least 967 mementos were working and displayed the label. Similarly for the VTR label, at least 213 new UI mementos out of 559 have a label in them. 

\begin{figure*}[h!]
\centerline{\includegraphics[width=0.8\textwidth]{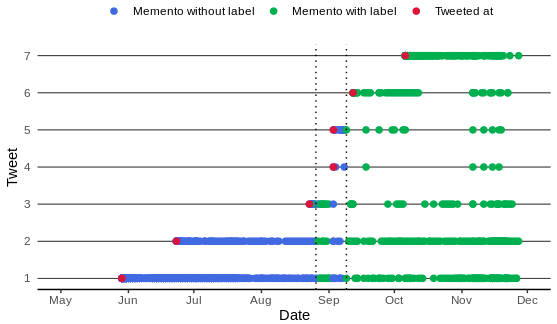}}
\caption{The old UI mementos of the seven earliest tweets that were labeled for violating Twitter rules. The red dot shows when each tweet was created, blue dot denotes mementos without labels, and green dot illustrates mementos with labels.}
\label{fig:VTR_timeline}
\end{figure*}

Even though labels were initially not a part of the old UI, VTR labels began appearing in old UI mementos in IA. We analyzed the presence of this label in the old UI mementos to determine when Twitter added this label to its old UI. 
We scraped the HTML of all the old UI mementos of the VTR labeled tweet IDs to determine which contain the label. Figure~\ref{fig:VTR_timeline} illustrates the old UI mementos of the seven earliest tweets that were labeled for violating Twitter rules. The mementos marked with blue do not contain the label, and the mementos marked with green contain the label.

The \ref{fig:VTR_timeline} shows that the label did not appear until August 26, 2020 (dotted line 1) in old UI mementos even for the tweet created on May 29, 2020 (Tweet 1). However, after September 9, 2020 (dotted line 2), labels can be seen in old UI mementos for all seven tweets. For the period between August 26, 2020 and September 9, 2020, we can see mementos both with and without labels. This shows that the change was not instantaneous and happened over approximately two weeks.

\section{Future Work}
We would like to extend our research to analyze other accounts that were suspended by Twitter following the US Capitol riot on January 6, 2021. The archived content will be the primary source for those accounts, since no other accounts are popular enough to have third party archives like thetrumparchive.com and factba.se. For those accounts, the web archive will be the only record of their existence.

\section{Conclusions}
We analyzed the difficulties that most web archives experience when archiving Twitter pages after Twitter stopped supporting its old UI in May 2020. Because of its popularity and historical significance, we used the @realDonaldTrump account to analyze the loss of information in web archives caused by Twitter’s UI change. We observed how the replay of a web page does not always match the live web expectations. We explored the differences between the old UI and the new UI. We also elaborated on how web archives are facing difficulties in archiving the new UI, and how old UI mementos are still more prominent (93.3\% of 1,324,363 mementos) than new UI mementos in the archives. 

The importance of the new UI is that it carries information that the old UI does not. We discussed how the ``Fact-check” labels are not a part of Twitter’s old UI. This indicates how there is a potential for losing valuable information if we are unable to archive and replay the new UI mementos properly. Even if we only consider mementos of labeled tweets, 96.69\% (191,221 out of 197,774) are of Twitter’s old UI.

Additionally, we also discussed how new UI mementos are vulnerable to significant temporal violations. For example, the tweets in a new UI memento could be coming from sometime between four days in the past or eight days in the future, resulting in replaying an account page memento that may have never existed on the live web. We found that 48.8\% of our 1,790 complete new UI mementos were temporally violative, i.e. they had at least one missing tweet relative to what a user would have seen on the live web at the time of archiving.
Information loss is inevitable when independently operating web archives try to capture the distributed web that is governed by different non-cooperating entities with different goals. Some of these losses (such as temporal violations) can be detected systematically or identified via curation. For better contextualization and transparency, web archives may acknowledge such issues at the time of replay. However, some information loss (such as missing Twitter labels) can be difficult to identify after the fact. These uncertainties in the replay allows for mis/disinformation in the web archives. However, the awareness of these problems can help historians and researchers not draw inaccurate conclusions while using the new UI mementos. 


\bibliographystyle{IEEEtran}
\bibliography{refs}

\begin{thebibliography}{10}
\providecommand{\url}[1]{#1}
\csname url@samestyle\endcsname
\providecommand{\newblock}{\relax}
\providecommand{\bibinfo}[2]{#2}
\providecommand{\BIBentrySTDinterwordspacing}{\spaceskip=0pt\relax}
\providecommand{\BIBentryALTinterwordstretchfactor}{4}
\providecommand{\BIBentryALTinterwordspacing}{\spaceskip=\fontdimen2\font plus
\BIBentryALTinterwordstretchfactor\fontdimen3\font minus
  \fontdimen4\font\relax}
\providecommand{\BIBforeignlanguage}[2]{{%
\expandafter\ifx\csname l@#1\endcsname\relax
\typeout{** WARNING: IEEEtran.bst: No hyphenation pattern has been}%
\typeout{** loaded for the language `#1'. Using the pattern for}%
\typeout{** the default language instead.}%
\else
\language=\csname l@#1\endcsname
\fi
#2}}
\providecommand{\BIBdecl}{\relax}
\BIBdecl

\bibitem{IntroducingNewTwitter:Twitter}
Twitter, ``{Introducing a new Twitter.com},''
  \url{https://blog.twitter.com/en_us/topics/product/2019/introducing-a-new-Twitter-dot-com.html},
  2019.

\bibitem{memento:rfc}
H.~{Van de Sompel}, M.~L. Nelson, and R.~Sanderson, ``{HTTP framework for
  time-based access to resource states -- Memento, Internet RFC 7089},''
  http://tools.ietf.org/html/rfc7089, 2013.

\bibitem{PermanentSuspensionofTrump:Twitter}
Twitter, ``{Permanent suspension of @realDonaldTrump},''
  \url{https://blog.twitter.com/en_us/topics/company/2020/suspension.html},
  2021.

\bibitem{TwitterWasAlreadyDifficult:KritikaHimarsha}
K.~Garg and H.~Jayanetti, ``{Twitter Was Already Difficult To Archive, Now It's
  Worse!}''
  \url{https://ws-dl.blogspot.com/2020/07/2020-07-15-twitter-was-already.html
  }, 2020.

\bibitem{UpdatingOurApproach:YoelNick}
Y.~Roth and N.~Pickles, ``{Updating our approach to misleading information},''
  \url{https://blog.twitter.com/en_us/topics/product/2020/updating-our-approach-to-misleading-information.html},
  2020.

\bibitem{ainsworth2014framework}
S.~G. Ainsworth, M.~L. Nelson, and H.~{Van de Sompel}, ``A framework for
  evaluation of composite memento temporal coherence,'' Tech. Rep.
  arXiv:1402.0928, 2014.

\bibitem{NewTwitterUI:HimarshaKritika}
H.~Jayanetti and K.~Garg, ``{New Twitter UI: Replaying Archived Twitter Pages
  That Never Existed},''
  \url{https://ws-dl.blogspot.com/2020/11/2020-11-04-new-twitter-ui-replaying.html},
  2020.

\bibitem{TweetJan6:Twitter}
{Twitter Safety}, ``{An update following the riots in Washington, DC},''
  \url{https://blog.twitter.com/en_us/topics/company/2021/protecting--the-conversation-following-the-riots-in-washington--},
  2021.

\bibitem{wells2020trump}
C.~Wells, D.~Shah, J.~Lukito, A.~Pelled, J.~C. Pevehouse, and J.~Yang, ``Trump,
  {Twitter}, and news media responsiveness: A media systems approach,''
  \emph{New Media \& Society}, vol.~22, no.~4, pp. 659--682, 2020.

\bibitem{pain2019president}
P.~Pain and G.~Masullo~Chen, ``The president is in: Public opinion and the
  presidential use of {Twitter},'' \emph{Social Media + Society}, vol.~5,
  no.~2, 2019.

\bibitem{ott2020twitter}
B.~L. Ott and G.~Dickinson, ``The {Twitter} presidency: How {Donald Trump's}
  tweets undermine democracy and threaten us all,'' \emph{Political Science
  Quarterly}, vol. 135, no.~4, pp. 607--636, 2020.

\bibitem{ExaminingTwitter}
K.~Starbird and C.~Miller, ``{Examining Twitter’s policy against
  election-related misinformation in action},''
  \url{https://www.eipartnership.net/policy-analysis/twitters-policy-election-misinfo-in-action},
  2020.

\bibitem{Response:Twitter}
{Twitter}, ``{Twitter API {HTTP} status codes},''
  \url{https://developer.twitter.com/en/support/twitter-api/error-troubleshooting},
  2020.

\bibitem{AdditionalSteps:VijayaKayvon}
V.~Gadde and K.~Beykpour, ``{Additional steps we're taking ahead of the 2020 US
  Election},''
  \url{https://blog.twitter.com/en_us/topics/company/2020/2020-election-changes.html},
  2020.

\bibitem{TwitterAddedLabels:HimarshaKritika}
K.~Garg and H.~Jayanetti, ``{Twitter Added Labels On Its Old User Interface},''
  \url{
  https://ws-dl.blogspot.com/2020/12/2020-12-08-twitter-added-labels-on-its.html},
  2020.

\bibitem{acker2020weaponization}
A.~Acker and M.~Chaiet, ``The weaponization of web archives: Data craft and
  {COVID-19} publics,'' \emph{Harvard Kennedy School (HKS) MisInformation
  Review}, 2020.

\bibitem{TwitterRewritesYourURLS:HimarshaKritika}
H.~Jayanetti and K.~Garg, ``{Twitter rewrites your URLs, but assumes you’ll
  never rewrite theirs: more problems replaying archived Twitter},''
  \url{https://ws-dl.blogspot.com/2021/01/2020-01-22-twitter-rewrites-your-urls.html},
  2021.

\bibitem{CookiesViolations:Sawood}
S.~Alam, ``{Cookie Violations Cause Archived Twitter Pages to Simultaneously
  Replay in Multiple Languages},''
  \url{https://ws-dl.blogspot.com/2019/03/2019-03-18-cookie-violations-cause.html},
  2019.

\bibitem{alam2019impact}
S.~Alam, P.~Vargas, M.~C. Weigle, and M.~L. Nelson, ``Impact of {HTTP} cookie
  violations in web archives,'' Tech. Rep. arXiv:1906.07141, 2019.

\bibitem{ainsworth2015only}
S.~G. Ainsworth, M.~L. Nelson, and H.~{Van de Sompel}, ``Only one out of five
  archived web pages existed as presented,'' in \emph{Proceedings of the 26th
  ACM Conference on Hypertext \& Social Media}, 2015, pp. 257--266.

\bibitem{CNNunarchivable:JohnBerlin}
J.~Berlin, ``{2017-01-20: CNN.com has been unarchivable since November 1st,
  2016},''
  \url{https://ws-dl.blogspot.com/2017/01/2017-01-20-cnncom-has-been-unarchivable},
  2017.

\bibitem{Archiveweb:iilya}
I.~Kreymer, ``{Introducing ArchiveWeb.page - Local High-Fidelity Web Archiving
  directly in your browser},''
  \url{https://webrecorder.net/2021/01/18/archiveweb-page-extension.html},
  2021.

\bibitem{alam2019bundle}
S.~Alam, M.~C. Weigle, M.~L. Nelson, M.~Klein, and H.~{Van de Sompel},
  ``Supporting web archiving via web packaging,'' Tech. Rep. arXiv:1906.07104,
  2019.

\bibitem{webrecorder:iilya}
I.~Kreymer, ``{Webrecorder: Developing an Open-Source High-Fidelity Web
  Archiving Toolset},''
  \url{https://2019.code4lib.org/talks/Webrecorder-Developing-an-OpenSource-HighFidelity-Web-Archiving-Toolset},
  2019.

\bibitem{Scalecollection:ArchiveIt}
K.-R. Blumenthal, ``{The stack: High fidelity web collecting at scale with
  Brozzler},'' \url{https://archive-it.org/blog/post/the-stack-brozzler/},
  2020.

\bibitem{Pywb2:iilya}
I.~Kreymer, ``{Pywb 2.0: technical overview and Q\&A},''
  \url{https://netpreserve.org/ga2018/workshops/pywb-2-0-technical-overview-and-qa/},
  2018.

\bibitem{LabelIntroduction:Twitter}
Twitter, ``{Updating our approach to misleading information},''
  \url{https://blog.twitter.com/en_us/topics/product/2020/updating-our-approach-to-misleading-information},
  2020.

\bibitem{alam2016memgator}
S.~Alam and M.~L. Nelson, ``{MemGator - A portable concurrent memento
  aggregator: Cross-platform {CLI} and server binaries in Go},'' 2016, pp.
  243--244.

\bibitem{CookiesAreWhy:SawoodPlinio}
S.~Alam and P.~Vargas, ``{Cookies Are Why Your Archived Twitter Page Is Not in
  English},''
  \url{https://ws-dl.blogspot.com/2018/03/2018-03-21-cookies-are-why-your.html},
  2018.

\bibitem{LinksMystery:dshr}
D.~S.~H. Rosenthal, ``The 47 links mystery,''
  \url{https://blog.dshr.org/2019/03/the-47-links-mystery.html}, 2019.

\bibitem{TrumpsTweets:Ed}
E.~Summers, ``{Trump's Tweets},''
  \url{https://inkdroid.org/2021/01/21/trumps-tweets/}, 2021.

\bibitem{SearchingWebArchives:MNS}
M.~N. Siddique, ``{Searching Web Archives for Unattributed Deleted Tweets From
  Politwoops},''
  \url{https://ws-dl.blogspot.com/2019/08/2019-08-03-searching-web-archives-for.html},
  2019.

\bibitem{TweetedAt:NaumanSawood}
M.~N. Siddique and S.~Alam, ``{TweetedAt: Finding Tweet Timestamps for Pre and
  Post Snowflake Tweet IDs},''
  \url{https://ws-dl.blogspot.com/2019/08/2019-08-03-tweetedat-finding-tweet.html},
  2019.

\bibitem{selenium}
Selenium, ``{Selenium Client Driver},''
  \url{selenium.dev/selenium/docs/api/py/}, 2018.

\bibitem{twitterlabels_2021}
K.~Garg and H.~Jayanetti, ``{TwitterLabels},''
  \url{https://github.com/oduwsdl/TwitterLabels}, 2021.

\bibitem{selenium-wire}
W.~Keeling, ``{Selenium Wire},'' \url{https://pypi.org/project/selenium-wire/},
  2020.

\end{thebibliography}

\end{document}